\documentclass[aps,preprint,nofootinbib,groupedaddress]{revtex4}
\usepackage[dvips]{graphicx}
\usepackage{amsmath,amssymb,graphicx,subfigure}

\newcommand{\be}{\begin{equation}}
\newcommand{\ee}{\end{equation}}
\newcommand{\bea}{\begin{eqnarray}}
\newcommand{\eea}{\end{eqnarray}}

\def\slashchar#1{\setbox0=\hbox{$#1$}           
   \dimen0=\wd0                                 
   \setbox1=\hbox{/} \dimen1=\wd1               
   \ifdim\dimen0>\dimen1                        
      \rlap{\hbox to \dimen0{\hfil/\hfil}}      
      #1                                        
   \else                                        
      \rlap{\hbox to \dimen1{\hfil$#1$\hfil}}   
      /                                         
         \fi}                                         %

\begin{document}

\title{The Scalar Glueball in the Instanton Vacuum}
\author{M.C. Tichy$^{1,2}$ and P. Faccioli$^{2,3}$\footnote{Corresponding author. E-mail:~faccioli@science.unitn.it}}

\affiliation{$^1$ Insitut f\"ur Theoretische Physik, Universit\"at T\"ubingen ,  Auf der Morgenstelle 14 , D-72076 T\"ubingen, Germany\\
$^2$ Dipartimento di Fisica , Universit\`a degli
Studi di Trento, Via Sommarive 15, Povo (Trento) 38050 Italy.\\
$^3$ I.N.F.N, Gruppo Collegato di Trento, Via Sommarive 15, Povo (Trento) 38050 Italy.}

\begin{abstract}

We study the contribution of instantons to the binding and the mass of the lightest scalar glueball, in gluondynamics. We show that the short-range correlations introduced by such  non-perturbative  vacuum fluctuations are sufficient to give raise to a  scalar glueball bound-state, with mass in good agreement with the results of recent lattice calculations.  
\end{abstract}

\maketitle

\section{Introduction}

The question whether the hadron spectrum contains glueballs and hybrids has received much attention, since the establishment of QCD.
From the experimental point of view, such exotic resonances are believed to be most effectively excited in photo-production experiments. However, the available data do not allow to unambiguously identify the gluonic exitations, among the ordinary ~$|q \bar q\rangle$ states \cite{pheno}. In order to help clarify the ambiguous identifications below $c \bar c$ threshold, dedicated spectroscopic experiments with linearly polarized photon beams will be performed by the {\it GlueX} Collaboration~\cite{gluex} at Jefferson Laboratory. The PANDA experiment at GSI \cite{Bettoni:2005ut} has been designed to discover glueballs up to masses of 4.5 GeV and will be fully operational by 2012. 

From the theoretical point of view, the problem of understanding the gluon content of hadronic wave-functions is deeply connected with the problem of identifying the dominant non-perturbative vacuum gauge configurations,  responsible for the  dynamical quark-gluon and gluon-gluon correlations, at intermediate and large distances. In particular, the spectrum of gluondynamics encodes the information about the vacuum gauge configurations responsible for the non perturbative gluon-gluon interaction.

Gluondynamics has been extensively investigated with quenched Lattice QCD simulations~\cite{liu,mayer}.
These studies have established that the lightest glueball is a $J^{PC}=0^{++}$ state, with mass in the range $1.500-1.750$~GeV.
In particular, Chen {\it et al.}\cite{liu} recently reported a mass of $M_{0^{++}}=1.710$~GeV, obtained with anisotropic lattices, while Meyer and Teper\cite{mayer} used much finer isotropic lattices, and reported a value 
$M_{0^{++}}= 1.475$~GeV. Both such calculations are affected by a $\sim~2\%$ statistical error and a $\sim 3\%$ systematic error, associated with the  determination of the physical lattice spacing.

It has also been established that the physical properties of this state  are remarkably different from those of all other glueballs. First of all, it is significantly lighter, since tensor and pseudo-scalar glueballs have masses  $M_{2^{++}}\simeq 1.4~M_{0^{++}}$ and $M_{0^{-+}}\simeq~1.5-1.8~M_{0^{++}}$, respectively.  In addition, the scalar glueball is an unusually compact object,  of size\footnote{The size of the glueball can be estimated, e.g. by measuring the exponential decay  of Bethe-Salpeter wave-function in Euclidean space} $\sim~0.2$~fm~\cite{sizeglue}, while the $2^{++}$ and $0^{-+}$  states have much larger sizes $\gtrsim 0.8$~fm, i.e. comparable with that of the typical hadrons. 
It is interesting to note that in the MIT Bag Model, in which all states are bound exclusively by confining forces, 
the scalar and tensor glueballs are predicted to have comparable masses of $\simeq 1~$GeV and comparable sizes of $\sim 0.8$~fm.

Given the very small size of the scalar glueball, one expects that the dominant forces responsible for its binding must be provided by short-sized vacuum fluctuations, of correlation length $\sim 0.2~\textrm{fm}$. Consequently, typical energy-momentum transfers inside a scalar glueball occur at the chiral symmetry breaking scale   $\Lambda_{\chi}\sim~$1~GeV, rather than at the confinement scale $\Lambda_{QCD}\sim~250$MeV.

It has long been argued that instantons of size $\sim~0.3$~fm represent the main vacuum gauge field configurations responsible for the  non-perturbative dynamics at the chiral scale~\cite{shuryak82, diakonov84}. Several recent lattice  studies~ \cite{gattringer, faccioli} have provided strong evidence in support of such an hypothesis. Hence, it is natural to raise the question whether the correlations generated by  such semi-classical vacuum fluctuations are sufficiently strong  to bind the lightest glueball state.

The role of instanton-induced interactions in light hadrons has been investigated in a number of papers. 
It was shown that they give raise to a nucleon and a pion with realistic masses~\cite{nucleonILM,mymasses} and electroweak structure~\cite{nucleonFF, pionFF, delta12}. In addition, the dependence of the mass of these hadrons on the quark mass agrees extremely well with the predictions of chiral perturbation theory and with the available results of lattice QCD simulations~\cite{nucleonFF}.
It was also shown that lowest-lying vector and axial-vector meson resonances can be bound by instanton-induced chiral forces. In these resonances, however,  instanton effects are much weaker than in the nucleon and pion, as they are suppressed by a power of the instanton ensemble's diluteness, $\kappa\sim~0.1$. As a result, the mass of the resonances turns out to be about $30\%$ heavier than the experimental value. This fact can be interpreted as a signature of the fact that other sources of non-perturbative correlations ---presumably confining forces---  begin to play an important role in these systems.
We argue that a similar situation is realized in the glueball sector: the lightest scalar is bound predominantly by instantons, while the structure of the spectrum of its resonances is presumably shaped by other gauge configurations responsible for color confinement. 

The instanton contribution to the lightest glueballs has been investigated in the framework of operator product expansion (OPE) \cite{earlyOPE1,earlyOPE2,forkel} or by computing directly point-to-point correlators at short and intermediated distances, in the Random Instanton Liquid Model  and in the Interacting Instanton Liquid Model \cite{shusha}. 
Such studies allowed to establish that, at least at the qualitative level, the ordering of the states in the light glueball
spectrum fits well in the instanton picture: in fact, the $\mathcal{O}(\kappa)$ instanton interaction is attractive in the scalar channel,  is suppressed in the tensor channel and  is  repulsive in the pseudo-scalar channel.
Remarkably, the same dynamical mechanism can explain the ordering of the lightest $q\bar q$ excitations. In fact, the $\mathcal{O}(\kappa)$ instanton correlations are attractive in the pion, are suppressed in the $\rho$-meson and are repulsive in
 the $\eta'$-meson. Finally, the same argument applies also to the lowest-lying sector of the light baryon spectrum, where leading instanton effects are attractive in the nucleon, but suppressed in the $\Delta$-isobar. It 
 should be stressed that the short-range interactions induced by instantons in these systems can at most provide binding, but do not confine.

At a  quantitative level, the most sophisticated instanton-improved sum-rule calculation available~\cite{forkel} leads to a scalar mass $m_S=1.25\pm 0.20$~GeV, while numerical simulations of point-to-point correlators in Interacting Instanton Liquid Model give $m_S\simeq1.5~$GeV~\cite{shusha}. 
The fairly good agreement between these calculations  may be interpreted as evidence in favor of the existence of scalar glueball in the instanton vacuum.
On the other hand, it should be stressed that both calculations rely on the assumption that the spectral function in the scalar channel contains a glueball bound-state, separated by a ``perturbative'' continuum. 
The existence of a pole in the Green's function is postulated, and its location is determined from a fit of the calculated correlators, in either  Borel or coordinate space.

In this work, we complement these studies using a different approach. We address the question of the existence of a scalar glueball without  {\it assuming} that the corresponding two-point correlation function contains a pole. 
To do so, we perform an analysis which is quite similar to the one used to extract the mass in lattice simulations. 
However, rather than looking at the long Euclidean time behavior of the zero-momentum projected correlation functions, we focus on finite 
momentum states and we perform a calculation in a finite range of Euclidean times.
This choice allows to reduce the model dependence associated to the unknown structure of the instanton-intanton correlations.
This way, we find evidence for a scalar glueball bound state, whose mass is in good agreement with the results of recent lattice calculations. 

The paper is organized as follows. In section \ref{correlators} we describe the method used to infer the existence of a glueball and to determine its mass.
In sections \ref{ilm} and \ref{SIAcalc} we present our calculation. Section \ref{considerations} contains some general considerations on the structure of gluonic correlations responsible 
for the binding of the glueball, in this model. Finally, the main results and conclusions are summarized in \ref{conclusions}. 

\section{Scalar Two-Point Correlation Function and the Momentum-Dependent Effective Mass}
\label{correlators}

In quantum field theory, the information about the spectrum is encoded in the two-point correlation functions. 
In our specific case,  we are interested in the scalar correlation function 
\be
\Pi_S(x)=\langle 0| T[S(x)~S(0)]|0\rangle, 
\label{p2p}
\ee
where $x=(\tau,{\bf x})$ is a space-time point in the Euclidean space and $S(x)$ is proportional to the action density operator 
\be
S(x)= g^2 G^a_{\mu \nu}(x)\,G^a_{\mu \nu}(x),  
\label{Sope}
\ee
which excites  states with $J^{P C}=0^{++}$ quantum numbers. 

Using the spectral representation, it is immediate to show that, in the limit of large Euclidean separation  $|x|\to \infty$, the point-to-point correlator (\ref{p2p}) converges to the square of the gluon condensate.
At  large-but-finite space-time distances, $\Pi_S(x)$ receives contribution also from the lowest-lying vacuum excitation, with  $0^{++}$ quantum numbers. 
Lattice simulations have shown that, in gluon-dynamics, the lightest vacuum excitation is a glueball. In  QCD, if a scalar glueball exists it must be unstable and decay into  a two-pion state.  In general, in order to extract
 information about the scalar glueball from the large-distance behavior of the correlation function (\ref{p2p}) one has to subtract  the gluon condensate 
contribution and assume that the purely gluonic operator $S(x)$ couples weakly to hadron excitations containing quarks.
 
For the purpose of identifying a signature of the existence of a glueball bound-state in the spectrum, it is  convenient to introduce the notion of effective energy. We consider the momentum-projected correlation function,
\be
G_S(\tau, {\bf p})= \int d^3 {\bf x} ~ e^{i {\bf p\cdot x}}~\Pi_S(\tau,{\bf x}).
\label{proj}
\ee

The effective energy $E_{eff}(\tau, {\bf p})$ is defined as the logarithmic derivative of $G_S(\tau,{\bf p})$,
\be
E_{eff}(\tau,{\bf p})= -\frac{d}{d\tau} \log G_S(\tau, {\bf p})
\ee

{\it If} the  spectrum contains a scalar glueball bound-state of mass $M_{0^{++}}$, then  at large Euclidean times the effective energy $E_{eff}(\tau,{\bf p})$ must become independent on $\tau$ and converge to the kinetic energy of such a state, propagating with momentum ${\bf p}$: 
\be
\lim_{\tau\to\infty} E_{eff}(\tau,{\bf p}) = \sqrt{{\bf p}^2 + M_{0^{++}}^2}.
\ee

In lattice QCD simulations it is convenient to project onto states with zero momentum, ${\bf p}=0$, since projections on finite momentum states are usually more noisy. This way, one can directly read-off the mass of the lowest state from the plateau in the effective-energy, at large Euclidean times. 

However, one price to pay for such a choice is that one needs to subtract the contribution of the gluon condensate from the scalar two-point correlator. If the fluctuations of the correlator are very large, or if the simulation box is not sufficiently large, the result of such a subtraction may very noisy or affected by systematic errors.
In this work we choose to adopt a more general standpoint and consider the projection of the point-to-point Green's function to some   finite momentum ${\bf p}$. Hence, we define a {\it momentum-dependent effective mass}, 
\be
M_{eff}(\tau,{\bf p}) = \sqrt{E^2_{eff}(\tau,{\bf p}) - {\bf p}^2}. 
\label{pMeff}
\ee
If the lowest scalar excitation in the spectrum is a single-particle bound-state, then in the large Euclidean time limit $M_{eff}(\tau,{\bf p})$ must stop depending on $\tau$ and on ${\bf p}$ and converge to the glueball's mass:
\bea
\lim_{\tau\to\infty} M_{eff}({\tau,\bf p}) =  M_{0^{++}}.
\eea

We conclude this section by discussing how the range of Euclidean times where the lowest excited state contribution becomes dominant depends on the projection momentum $\bf p$. 
The spectral reppresentation of the momentum-projected correlator (\ref{proj}) leads to the expression
$$ G(\tau,{\bf p}) = \sum_n c_n({\bf p}) e^{-E_n({\bf p})\tau }, $$
where $E_n(\bf p)$ is the energy of the n-th eigenstate of momentum ${\bf p}$ and $c_n({\bf p})$ is the corresponding overlap with the scalar operator.

The momentum dependence of the couplings $c_n(p)$ is completely determined by Lorentz symmetry. For example, the coupling of the scalar operator to the scalar glueball state reads
\be \langle 0| S(0)|G_{ O^{++}}({\bf p})\rangle =\frac{M_{0^{++}}}{\sqrt{{\bf p}^2 + M_{0^{++}}^2}}\langle 0| S(0)| G_{O^{++}}({\bf 0})\rangle. \ee

Let us focus on large Euclidean times for which only the two lowest vacuum excitations contribute. 
If these are single particle states, in the zero-momentum projection we have $E_1=m_1$ and $E_2=m_2$. Hence, the time required to filter the lowest state scales with the inverse mass gap $ \tau \sim \frac 1 {m_2-m_1}$. 

At large momentum the filtering of the lowest state becomes more difficult, since 
$E_1=\sqrt{m_1^2+{\bf p}^2} \approx E_2=\sqrt{m_2^2+{\bf p}^2}$. Consequently, the contributions of the excitations require longer times to die out.

\begin{figure}[t]
  \begin{center}
  \vspace{5mm}
\includegraphics[width=5cm,angle=270]{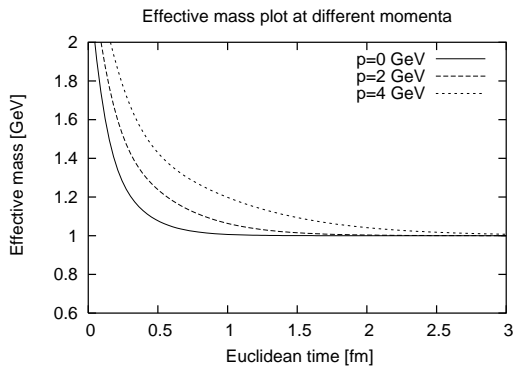}
  \end{center}
\caption{Structure of the momentum-dependent effective mass plot  at different momenta, for a spectrum consisting three boundstates of mass 1 GeV, 2 GeV  and 4 GeV.}
\label{XX}
\end{figure}

We now  estimate the dependence of such a time scale on the momentum ${\bf p}$. Let $G_1(\tau,p)$ and $G_2(\tau,p)$ be the contribution of the lowest two excitations to the momentum projected correlator, i.e.\footnote{Note that we are neglecting the non-exponential dependence of the coefficients $c_1$ and $c_2$ on the momentum ${\bf p}$. }
$$ G(\tau,{\bf p})=c_1({\bf p}) e^{-E_1({\bf p})\tau} + c_2({\bf p}) e^{-E_2({\bf p})\tau} =: G_1(\tau,{\bf p})+ G_2(\tau,{\bf p}). $$ The lowest excitation dominates if  $$ G_1(\tau,{\bf p}) \ge f \cdot G_2(\tau,{\bf p}), $$ where $f$ is some large parameter, $f \gg 1$.  This will be the case for times larger than some given time $\tau_0$, defined as  $$ \tau_0 = \frac{\log(f \frac{c_2}{c_1})}{E_2({\bf p})-E_1({\bf p})}.$$ For ultra-relativistic momenta $p \gg m_1,m_2$, this leads to  $ \tau_0 \propto p.$
In Fig. \ref{XX}, we show the expected behaviour of the corresponding effective mass at different momenta, when several boundstates are present in the spectrum. One can clearly see that at higher momenta, the plateau which can be associated to the lowest state is shifted to larger Euclidean times.

\section{Scalar Two-Point Correlator in the Instanton Model}
\label{ilm}

The classical field of an individual instanton or antiinstanton is specified by a set of collective coordinate $\Omega= (z,\rho, U)$ where $z$ is the position of the pseudo-particle, $\rho$ is its size and $U$ is a SU(3) matrix which selects the orientation in color space.

Let $A_\mu^{cl}(x;\Omega_1,....,\Omega_{N=N_++N_-})$ be a  back-ground  field, constructed  from the classical fields of $N_+$ instantons and $N_-$ anti-instantons, according to some prescription. 
For example, the simplest choice is the so-called sum {\it ansatz}, in which amounts to adding up the contribution of all individual fields: 
\be
A^{cl.}_\mu(x)= \sum_{i=1}^{N}~A^i_\mu(x;\Omega_i),
\ee
where $A^i_{\mu}(x;\Omega_i)$ is the classical field of the $i$-th instanton.

The Instanton Model for gluon-dynamics is defined by replacing the path-integral over the gauge field configurations by a sum over the configurations of a grand-canonical statistical ensemble of singular gauge instantons and antiinstantons:
\begin{equation}
\label{eq:partf}
\mathcal{Z}_{ILM} =\sum_{N_+, N_-} \frac{1}{N_+!N_-!}\int\prod_i^{N} d\Omega_i
~ e^{-S_\textrm{YM}[A_\mu^{cl.}(\Omega_1,\ldots\Omega_N) ]}~e^{-F(\Omega_1,\ldots\Omega_N) }
\end{equation}
$S_{YM}[A_\mu^{cl.}(\Omega_1,...\Omega_N)]$ is the classical Yang-Mills action, evaluated in the classical background field 
$A^{cl.}_\mu(\Omega_1,...,\Omega_N)$. 
$F(\Omega_1,\dots, \Omega_N)$  is a function which formally represents the result of performing the functional integral over the fluctuations around the background fields, in the original path integral. 
Based on  the analogy with the classical statistical mechanics, this term  can be interpreted as the ``potential of mean-force'' between the pseudo-particles, due to  all the  degrees of freedom in the original system, which have been integrated out (i.e. the quantum gauge field configurations). Clearly, if one could  calculate  $F(\Omega_1,...,\Omega_N)$ from first principles, then Eq.(\ref{eq:partf})  would represent an exact  re-parametrization of the path-integral of gluon-dynamics, which would not stand on the semi-classical approximation. 

In general, both the classical action $S_{YM}[A_\mu^{cl}]$ and the quantum effective action  $F(\Omega_1,...,\Omega_N)$ introduce many-body correlations between the pseudo-particles.
If the ensemble is sufficiently dilute, it is possible to neglect three- and higher- body correlations and re-write the sum as:
\be
S_{YM}(\Omega_1,\ldots,\Omega_N) + F(\Omega_1,...,\Omega_N)\simeq \sum_i G_1(\Omega_i) + \sum_{i<j} G_2(\Omega_i, \Omega_j).
\label{G1G2}
\ee
where the functions $G_1$ and $G_2$ completely specify the model. In the appendix, we illustrate the choices of $G_1$ and $G_2$ which lead to the Interacting Instanton Liquid Model.


Let us now discuss the evaluation of the two-point scalar correlator (\ref{proj}),  in the Instanton Liquid Model. 
The first step consists in writing the gluon field  operator in the operators as a sum of a classical instanton background and a fluctuation field, 
\be
A_\mu(x)=A_\mu^{\textrm inst.}(x)+ A_\mu^{\textrm fluct.}(x).
\ee
At  distances much smaller than the typical instanton size,  the scalar correlator is dominated by the contribution of the fluctuation field, since the instanton contribution is suppressed by the finite size of the pseudo-particles. On the other hand,   at large distances, the correlator is  expected to be dominated by the background fields, which are non-perturbative. 
We then approximate the complete scalar correlator as a sum of  two terms: a short-distance part, estimated  from  {\it free} gluon propagation, and a long-distance, generated by the classical instanton fields~\cite{shuryakrev}:
\be
\Pi_S(x)=  \Pi_S^{\textrm{free}}(x) + \Pi_S^{\textrm{inst.}}(x),
\ee 
where the short-distance part is readily computed and gives
\be
\Pi_S^{\textrm{free}}(x) = - \frac{384 g^2}{\pi^4 x^8}
\label{PIfree}
\ee
and $g$ is the running coupling constant. 

 $\Pi_S^{\textrm{inst.}}(x)$ is the correlator in the classical multi-instanton background field, is evaluated in the canonical ensemble of $N_+$ instantons and $N_-$ anti-instantons and reads
\bea
\Pi_S^{\textrm{inst.}}(x)= \frac{ 1}{ Z^c_{ILM}}~\frac{1}{N_+!}\frac{1}{N_-!} ~\int \prod_{j=1}^{N} \mbox{d} \Omega_j  
e^{- \sum_k G_1(\Omega_k)- \sum_{k<l} G_2(\Omega_k,\Omega_k)}
~ S(x,A_\mu^{cl}) ~S(0,A_\mu^{cl}),
\label{fullILM}
\eea
where $S(x,A_\mu^{cl})$ is the scalar operator (\ref{Sope}) evaluated in the classical back-ground field, and $Z_{ILM}^c$ is the canonical partition function, 
\begin{equation}
 Z^c_{ILM} = \frac{1}{N_+!}\frac{1}{N_-!} ~\int \prod_{j=1}^{N} \mbox{d} \Omega_j  
 e^{-\sum_k G_1(\Omega_k)- \sum_{k<l} G_2(\Omega_k,\Omega_k) }
\label{Zc}
\ee

We now undertake one more approximation and assume that the scalar operator (\ref{Sope}) can be written as the sum of the contribution coming from the individual pseudoparticles:
\be
S(x; \Omega_1,\ldots,\Omega_N)= \sum_k s(x;\Omega_k),
\label{Ssum}
\ee
where $s(x;\Omega_k)$ is the action density evaluated at $x$, in the field of the $k-th$ speudo-particle and  reads,
\be
s(x;\Omega_k=(z_k,\rho_k,U_k)) =   \frac{192 \rho_k^4 }{\left(( x-z_k)^2+ \rho_k^2 \right)^4}.
\ee
Note that this approximation is motivated by the fact that the action density decreases very rapidly away from the instanton ---  according to a $1/x^8$ power-law---. Hence, the ansatz (\ref{Ssum}) becomes accurate if the system is dilute.

With such a choice, the instanton contribution (\ref{fullILM}) can be decomposed in two parts
\bea
\Pi_S^{\textrm{inst.}}(x)&=& \frac{ 1}{ Z^c_{ILM}}~\frac{1}{N_+!}\frac{1}{N_-!} ~\int \prod_{j=1}^{N} \mbox{d} \Omega_j  
e^{- \sum_k G_1(\Omega_k)- \sum_{k<l} G_2(\Omega_k,\Omega_k)}\nonumber\\
&\times&~\sum_{l, m}  s(x;\Omega_l) ~s(0;\Omega_m) \equiv \Pi_1(x) + \Pi_2(x)
\label{fullILM2}
\eea
 $\Pi_1(x)$ represents the contribution coming from the $N$ elements in the double-sum, in which the action density in $x$ and  $0$ is evaluated in the field of the same instanton. One finds
\be
\Pi_1(x) = N \frac{\int  d \Omega_1  
d_1(\Omega_1)~ s(\Omega_1,x) ~s(\Omega_1, 0)}{\int  d \Omega_1  d(\Omega_1)}, 
\label{fullILM3}
\ee
where
\be
d_1(\Omega_1) =  e^{- G_1(\Omega_1)}~ \int \prod_{j>1}^{N} \mbox{d} \Omega_j  
 e^{-\sum_{l>1}G_1(\Omega_l) +  G_2(\Omega_1,\Omega_l)}.
\ee 

The one-instanton distribution $d(\Omega)$ cannot in general be computed explicitly from first principles, because of the breakdown of the perturbative expansion at large instanton sizes \cite{thooft}. 
However, some of its general properties can be derived directly from symmetry considerations. In fact,  translational invariance imply that  $d(\Omega)$ does not depend on the instanton position $z$.  Moreover, gauge invariance of the action implies that does not depend on the color orientation. On the other hand, scale invariance is broken by the trace anomaly,  hence   $d(\Omega)$ represents the  instanton size distribution.

In conclusion, the $\Pi_1$ contribution to the point-to-point correlation function can be expressed as
\bea
\Pi_1(x) &=& \frac{N}{V}~ \frac{\int  d^4 z \int d \rho   
~d_1(\rho)~ s(\rho,x-z) ~s(\rho, -z)}{\int  d\rho~ d(\rho)}\nonumber\\
&=&    (192)^2  \,\frac{\bar{n}}{ \int d \rho ~d(\rho)}\, 
\int  d^4 z ~\int d \rho~ d_1(\rho)   \frac{ \rho^8  }{\left((x-z)^2 +  \rho^2 \right)^4}~ \frac{1}{\left(z^2 + \rho^2 \right)^4},
\label{fullILM4}
\eea
where $\bar n$ is the instanton density, which we shall not need to specify.

 Let us now consider the second contribution $\Pi_2(x)$, which comes from all terms in the double sum in   (\ref{fullILM}), in which the scalar operator in $x$ and in $0$ is evaluated in the field of different instantons.
\bea
\Pi_2^{\textrm{inst.}}(x)&=& N (N-1) \frac{\int \mbox{d} \Omega_1 \mbox{d} \Omega_2  ~d_2(\Omega_1,\Omega_2)
~ s(x;\Omega_1) ~s(0;\Omega_2)}{\int \mbox{d} \Omega_1 \mbox{d} \Omega_2 ~ d(\Omega_1,\Omega_2),
}\nonumber\\
&=& N (N-1) \frac{ \int \mbox{d}^4 z_1 d \rho_1 d U_1 \mbox{d}^4 z_2 d \rho_2 d U_2 ~d(z_1,\rho_1,U_1,z_2,\rho_2,U_2)
\frac{192 \rho_1^4}{((x-z_1)^2+\rho_1^2)^2} ~\frac{192 \rho_2^4}{((-z_2)^2+\rho_2^2)^2} }{\int \mbox{d}^4 z_1 d \rho_1 d U_1 \mbox{d}^4 z_2 d \rho_2 d U_2 ~d_2(z_1,\rho_1,U_1,z_2,\rho_2,U_2)},\nonumber\\
\label{fullILM5}
\eea
where
\be
d_2(\Omega_1,\Omega_2)= e^{- G_1(\Omega_1)-G_1(\Omega_2)-G_2(\Omega_1,\Omega_2)}~ \int \prod_{j>2}^{N} \mbox{d} \Omega_j  
 e^{-\sum_{l>2}G_1(\Omega_l) +  G_2(\Omega_1,\Omega_l)+G_2(\Omega_2,\Omega_l)}
\ee
is the two-body instanton density.

Some of the general features of the two-body density  $d_2(\Omega_1, \Omega_2)$  can be inferred from  symmetry arguments. 
First of all, we note that the dependence on the relative color orientation of the instantons is in general non-trivial. However, such dependence does not affect the calculation of the action density correlator, because the two terms $s(x;\Omega_1)$ and $s(0;\Omega_2)$ are separately gauge invariant.  Consequently the integral over the relative color orientation leads to an overall constant factor, which is canceled by the denominator. In addition, translational invariance implies that the density does not depend on $z_1$ and $z_2$ separately, but only on the relative distance, $z_2-z_1$.
Finally, we observe that infinitely separated instantons must be uncorrelated, so that the density correlation must approach a constant. 
Based on such considerations  $\Pi_2$ can be written as
\bea
\Pi_2(x) &=& N\,(N-1) \frac{ \int \mbox{d}^4 z_1 d \rho_1  \mbox{d}^4 z_2 d \rho_2 ~d'_2(z_1-z_2,\rho_1,\rho_2)
\frac{192 \rho_1^4}{((x-z_1)^2+\rho_1^2)^2} ~\frac{192 \rho_2^4}{(z_2^2+\rho_2^2)^2} }{\int \mbox{d}^4 z_1 d \rho_1  \mbox{d}^4 z_2 d \rho_2  ~d'(z_1-z_2,\rho_1,\rho_2)},
\eea
where 
\be
d_2'(|z_1-z_2|, \rho_1,\rho_2)= \int d U_1 d U_2 d(\Omega_1,\Omega_2).
\label{ddef}
\ee

It is immediate to verify by inspection that  the integrand is largest in the region in which $|z_1-z_2|\sim |x|$.
Hence, in the  large $|x|$ limit, in which the two-body density becomes constant, the contribution $\Pi_2(x)$ converges to 
\bea
\Pi_2(x) \stackrel {|x|\to\infty}{\rightarrow}  \frac{N(N-1)}{V} \times (32 \pi)^2.
\eea
Hence, in the thermodynamical limit $N,V\to\infty$, we obtain a prediction for the square of the gluon condensate.

In order to complete our calculation of the scalar correlator, we are left with specifying the one- and two-body density functions $d_1(\rho)$ and $d'_2(|z_1-z_2|, \rho_1,\rho_2)$.
The instanton density $d_1(\rho)$ has been evaluated on the lattice, using different methods (see e.g. \cite{latticerho} and references therein). On the other hand, no lattice calculation for $d'_2(|z_1-z_2|, \rho_1,\rho_2)$ has been performed so far. 
A possibility is to compute such a quantity in a model-dependent manner, for example using the choices for $G_1$ and $G_2$ of the  Interacting Instanton Liquid Model \cite{iilmfirst}, which is briefly described in the appendix. 
In this work, however, we choose to avoid introducing additional model dependence and  we try a different strategy, based on the Single Instanton Approximation (SIA)~\cite{sia}. 
The main idea is to  evaluate the momentum-dependent effective mass in a regime of momenta ${\bf p}$ and Euclidean time $\tau$, in which the scalar correlator is dominated by $\Pi_1(x)$, hence it is sensitive  mainly to the lattice-calculable 
function $d_1(\rho)$ and, not to  the unknown two-body density  $d'_2(|z_1-z_2|, \rho_1,\rho_2)$.

\section{Scalar Glueball in the Single Instanton Approximation}
\label{SIAcalc}

The SIA was first developed in \cite{sia, mymasses} and applied in \cite{nucleonFF, pionFF} to the calculation of electro-magnetic form factors. The starting point of this approach is to observe that, for sufficiently short-distances $x$, the scalar two-point  correlation function $\Pi(x)$ is saturated by the single-instanton term $\Pi_1(x)$ ---~which depends on the lattice calculable distribution $d_1(\rho)$---,
  and  becomes insensitive on the multi-instanton term $\Pi_2(x)$ --- which depends on  the unknown two-body density $d'_2(\Omega,\Omega)$---. 

The evaluation of the scalar glueball mass requires computing the momentum-projected correlation function (\ref{proj}) at large values of the Euclidean time $\tau$, where the contribution of the lowest bound-state can be isolated form from that of the higher scalar excitations.  Clearly, as the Euclidean time comparable with the typical distance between two pseudo-particles $\sim \bar{n}^{-1/3}$,  we expect the point-to-point correlator $\Pi(x)$ ---and therefore the momentum-projected correlator $G_S(\tau,{\bf p})$---  to start receiving significant contributions from the two-body term  $\Pi_2(x)$. 

The main idea of the present SIA calculation resides in observing that the range of Euclidean times in which the $G_S({\bf p}, \tau)$ is insensitive on the multi-instanton contribution $\Pi_2(x)$ can be increased,  
by projecting on states with large momenta ${\bf p}$. 

With such a choice, one suppresses the dependence on the multi instanton density $d'_2(|z_1-z_2|, \rho_1,\rho_2)$ which is responsible for the square of the gluon condensate. In addition,  the momentum projection integral
 in (\ref{proj}) becomes sensitive only to the value of the point-to-point correlator $\Pi({\bf x}, \tau)$ in the spatial points at a distance $\lesssim 1/|{\bf p}|$ from the Euclidean time axis.  Hence, the effects of correlations with pseudo-particles which are at a distance larger than $\sim 1/|{\bf p}|$ from such an axis is suppressed. 

The question we want to address in this section is whether there exist a window in ${\bf p}$ and $\tau$, in which the single-instanton term 
$\Pi_1(x)$ still dominates over the multi-instanton term $\Pi_2(x)$,  and the lowest scalar state dominates over its higher excitations. If this condition is verified, one should observe a range of Euclidean times in which the momentum-dependent effective mass $M_{eff} ({\bf p}, \tau)$ defined in Eq. (\ref{pMeff}) is independent on both ${\bf p}$ and $\tau$. As it was argued in sect.~\ref{correlators}, this would represent a clean signature of the existence of a scalar glueball bound-state in the instanton vacuum.

Based on such considerations, we now compute the momentum projection correlator $G_S({\bf p}, \tau)$  keeping only the one-instanton term $\Pi_1(x)$.  A number of independent lattice simulations have shown  that the instanton distribution $d_1(\rho)$ has a peak in the range $\rho\simeq 0.35-0.45$~fm, in good agreement with the estimate $\bar \rho\simeq 1/3$~fm, proposed long 
ago by Shuryak, based on phenomenological arguments \cite{shuryak82}.  However, the details of the size distributions are quite sensitive to the prescription used to extract the instanton content from a generic vacuum gauge configuration.  Hence, in this work, we choose to present the results obtained adopting the simplest form,
\be
d_1(\rho)= \bar n~ \delta(\rho-\bar \rho),
\label{simple}
\ee
where $\bar n\simeq 3~\textrm{fm}^{-4}$ is the instanton density, which was extracted from Monte Carlo simulations in the Interacting Instanton Liquid Model~\cite{nucleonILM}.
We have checked that the results obtained using a smeared size distribution which interpolates one obtained in lattice calculations agree within $10\%$ with those obtained using the simple parameterization (\ref{simple}). 
Such differences are arguably smaller than the accuracy of the present calculation. 
 
After integrating over the instanton-size, the single-instanton contribution to the scalar  point-to-point correlator reads~\cite{SIAglue,shusha}
\begin{eqnarray}
\Pi_S^{\textrm{inst.}}(x) =  (192)^2~\bar n \bar \rho^8 \int \mbox{d}^4 z  \frac{1  }{\left((x-z)^2 + \bar \rho^2 \right)^4}~ \frac{1}{\left(z^2 + \bar\rho^2 \right)^4}. 
\label{SIAcoord}
\end{eqnarray}

In order to construct the momentum-dependent effective mass (\ref{pMeff}), we take the spatial Fourier-transform 
of the calculated point-to-point correlator. The free contribution reads:
\bea
G_S^{\textrm{free}}(\tau,|{\bf p}|)= \frac{16\,g^{2}}{\tau^5}~e^{-\tau |{\bf p}|}(3+3\tau |{\bf p}| + \tau^2 {\bf p}^2) ,
\label{freecorr}
\eea
where the QCD coupling constant $g$ has been determined  at the chiral symmetry breaking scale, where the strong fine-structure constant is $\alpha_s\simeq~0.3$.
The instanton contribution in the SIA is
\begin{eqnarray}
G_S^{\textrm{SIA}}(\tau,{\bf |{\bf p}|}) &=&  64 \pi^4  \bar{\rho}^8 \int \mbox{d} z_4 \frac{e^{-|{\bf p}|(\sqrt{z_4^2+\bar{\rho}^2}+\sqrt{(\tau-z_4)^2+\bar{\rho}^2})} }{\left( (z_4^2+\bar{\rho}^2)((\tau-z_4)^2+\bar{\rho}^2)\right)^{5/2}} 
\cdot \left(3 + 3 \sqrt{z_4^2 + \bar{\rho}^2}|{\bf p}| + {\bf p}^2 (z_4^2+\bar{\rho}^2)\right)\nonumber\\&\cdot&~\left(3 + 3 \sqrt{(z_4-\tau)^2 + \bar{\rho}^2}|{\bf p}| + {\bf p}^2 ((z_4-\tau)^2+\bar{\rho}^2)\right).
\label{siacorr}
\end{eqnarray}
\label{results}
\label{siacalc}
\begin{figure}[t]
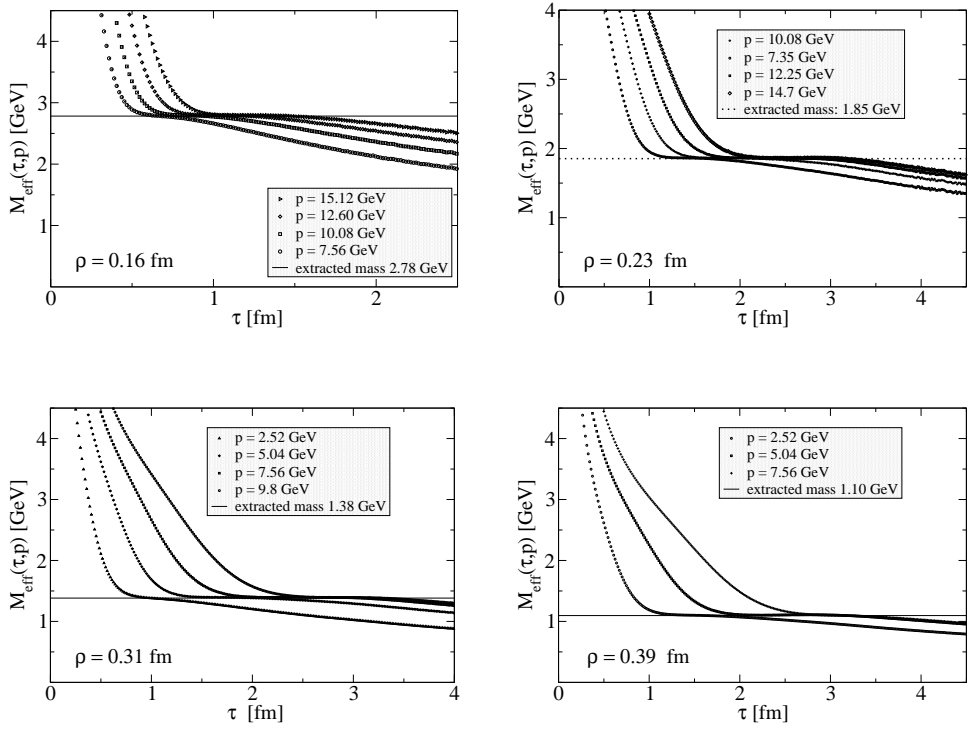

  \begin{center}
\includegraphics[width=6cm,angle=0]{plot-cutnumber02.eps}  \hspace{5mm}
\includegraphics[width=6cm,angle=0]{plot-cutnumber04.eps}\hspace{5mm}\\ \vspace{10mm}
\includegraphics[width=6cm,angle=0]{plot-cutnumber06.eps} \hspace{5mm}
\includegraphics[width=6cm,angle=0]{plot-cutnumber08.eps}
 \end{center}
 \caption{Effective-mass plots evaluated at fixed instanton-size for different momenta.}
\label{resultsSIA1}
\end{figure}

\begin{figure}[t]
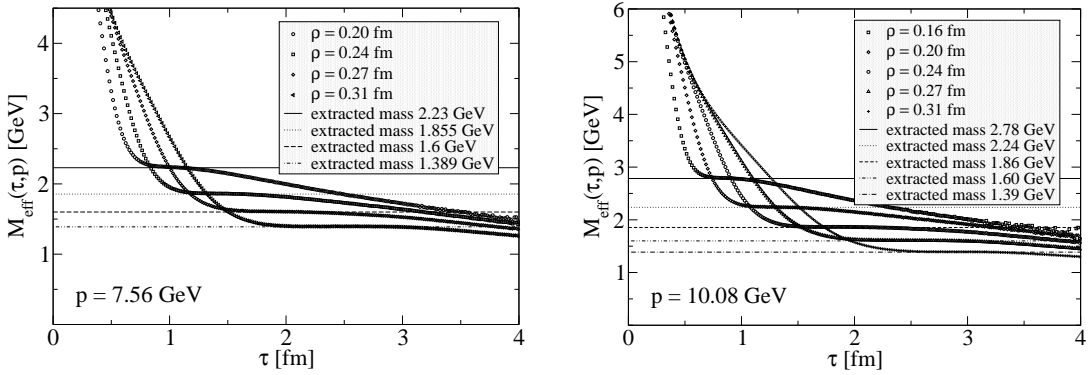

  \begin{center}
\vspace{1cm}
\includegraphics[width=6.9cm,angle=0]{plot-pp30.eps}  \hspace{5mm}
\includegraphics[width=6.7cm,angle=0]{plot-pp40.eps}\hspace{5mm} 
\end{center}
\caption{Effective-mass plots evaluated at fixed momentum, for different instanton-size.}
\label{resultsSIA2}
\end{figure}

From the momentum-projected correlation functions (\ref{freecorr}) and (\ref{siacorr}) it is immediate to construct the momentum-dependent effective mass:
\begin{equation}
M_{eff}(\tau,{\bf p}) =\sqrt{ \left(- \frac{\frac{\partial}{\partial t} \left( G_S^{\textrm{free}}(\tau,{\bf p}) +  G_{S}^{\textrm{SIA}}(\tau,{\bf p})\right) }{ G_S^{\textrm{free}}(\tau,{\bf p}) + G_{S}^{\textrm{SIA}}(\tau,{\bf p})}\right)^2 - {\bf p}^2 }.
\end{equation}

\label{discussion}
\begin{figure}[t]
  \begin{center}
  \vspace{5mm}
\includegraphics[width=8cm,angle=0]{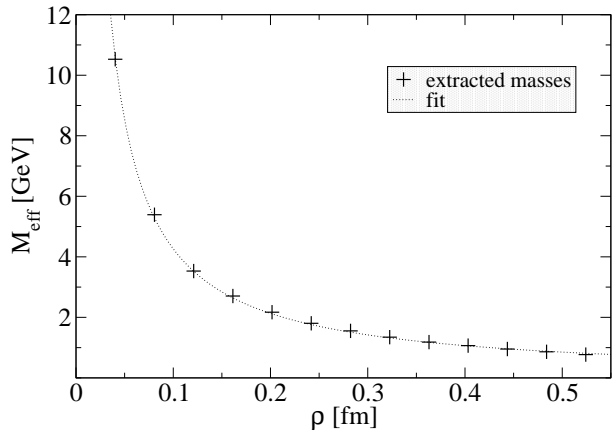}
  \end{center}
\caption{Scalar glueball mass calculated in the SIA, as a function of the instanton size $\bar \rho$.}
\label{Mrho}
\end{figure}

The results for  
$M_{eff}(\tau,{\bf p})$, 
calculated at different instanton sizes and different momenta are reported in 
Fig.~\ref{resultsSIA1} and Fig.~\ref{resultsSIA2}. 
Fig.~\ref{resultsSIA1} presents the momentum-dependent effective mass, in a range of momenta, for few fixed instanton sizes. On the other hand, Fig.~\ref{resultsSIA1} shows the effective mass, in a range of instanton sizes, for few fixed momenta. 
These plots clearly show that there exists a range of Euclidean times in which the momentum-dependent effective mass becomes independent on both momentum and Euclidean time. As was argued in sect.~\ref{correlators}, 
this represents a  signature of the existence of a scalar glueball bound-state in the instanton vacuum. 

On the other hand, at  large Euclidean times and for very small momenta, 
we observe that the calculated effective mass starts depending on $\tau$ and ${\bf p}$ again.  
This is well understood, since in such regimes the contribution of a single instanton to the correlation function is no longer dominant and therefore the SIA breaks down.
However, by stationary-phase analysis, we expect that if the value of the momentum ${\bf p}$ is increased, the integral (\ref{proj}) should become sensitive to shorter-sized point-to-point correlation functions. As a consequence, the range of validity of the SIA should extend for longer Euclidean times.

It should be stressed that, as the Euclidean time grows, we expect instanton-instanton correlations encoded in the contribution $\Pi_2(x)$ to the two-point correlation function to become important. Since this part is neglected, the plateau in the analytic effective mass plots shown above eventually breaks down, at large Euclidean times.

The complete list of masses extracted for different instanton sizes and different momenta are reported in Table~I.  
We stress the fact that, for any given fixed instanton size,  calculations performed at different momenta ${\bf p}$ lead to the {\it same} glueball mass. This represents an important consistency check of our calculation.

In order to further assess the accuracy of our calculation, we have analyzed the dependence of the glueball mass on the the  instanton size. Since the average instanton size is the only independent dimensional parameter in our model, the mass of the resulting scalar glueball should scale as $1/\bar \rho$. 
The plot in Fig.(\ref{Mrho}) shows that all our results can be fitted with very high accuracy by the function 
\be
M_{0^{++}}(\bar\rho)= \frac {2.16} {\bar\rho}.
\ee

Our final prediction for the scalar glueball mass is obtained by choosing the instanton size in the range $\rho\simeq 0.30-0.33$~fm, i.e. the value suggested by phenomenology and by lattice simulations. We find mass of 
$M^{SIA}_{0^{++}}\simeq 1.290-1.420$~GeV. Within its uncertainty, this result is consistent with the lattice result reported by Meyer and Teper, $M^{latt.}_{0^{++}}=1475(30)_{stat.}(65)_{sys.}$~MeV. 
Such an agreement is quite remarkable, given the simplicity of our calculation. It suggests that the binding of the compact scalar glueball mass is not strongly affected by confinement.  
We also note that our calculation agrees with the instanton-improved OPE prediction of Forkel $M^{OPE}_{0^{++}}=1.25(0.20)$~GeV.
\begin{table}[t]
	\centering
\begin{tabular}{|c||c|c|c|c|c|c|c|c|}\hline
& $|{\bf p}| =2.45  $ & $ |{\bf p}| =4.9$ & $|{\bf p}|=7.35$ & $|{\bf p}|=9.8$ & $|{\bf p}|=12.25$  & $|{\bf p}|=14.7$ & $|{\bf p}|=17.15$  & av. (err.)\\ \hline \hline
$ \bar{\rho}=0.04$ &$ - $&$  - $ &$  -  $&$  -   $&$  -  $&$  -  $&$10.528$ & $10.528~(-)$\\ \hline
$ \bar{\rho}=0.08$ &$ - $&$  - $ &$  -  $&$  -   $& 5.390 & 5.392 & 5.395& 5.392 (0.001) \\ \hline
$ \bar{\rho}=0.12$ &$ -$&3.430& 3.535& 3.550& 3.553 & 3.555& 3.555& 3.530 (0.020) \\ \hline
$ \bar{\rho}=0.16$ &$ - $&2.707& 2.695& 2.707& 2.707& 2.710& 2.705& 2.705 (0.002)\\ \hline
$ \bar{\rho}=0.20$  &2.166 & 2.166&  2.168& 2.171& 2.173& 2.171 & 2.173& 2.171  (0.001) \\ \hline
$ \bar{\rho}=0.24$ &1.801 & 1.798& 1.803& 1.803& 1.803& 1.806& 1.801&1.803 (0.001) \\ \hline
$ \bar{\rho}=0.28$ &1.534& 1.548& 1.556 & 1.556 & 1.556&  1.566 &$-$ &1.553 (0.005) \\ \hline
$ \bar{\rho}=0.32 $& 1.333 &  1.348 & 1.350 & 1.348 & $ -  $&$ -  $&$ -  $ &$1.345 (0.002)$ \\ \hline
$ \bar{\rho}=0.36$ & 1.186 & 1.188& 1.181 & 1.174 & $-$  &$ -  $&$ -  $ &$ 1.183 (0.002)$ \\ \hline
$ \bar{\rho}=0.40$ & 1.068 & 1.073 & 1.058 &$ -  $&$ -  $&$ -  $&$ -  $ & $1.066 (0.002)$\\ \hline
$ \bar{\rho}=0.44$ &$ 0.953$&$ 0.956  $&$-    $&$ -  $&$ -  $&$ -  $&$ -  $ & $ 0.954 (0.001)$\\ \hline
$ \bar{\rho}=0.48$ &$ 0.875$&$ 0.848  $&$ -   $&$ -  $&$ -  $&$ -  $&$ -  $ & $0.862 (0.010)$\\ \hline
$ \bar{\rho}=0.52$ &$ 0.786$&$ 0.750  $&$ -   $&$ -  $&$ -  $&$ -  $&$ -  $ & $0.769 (0.012)$\\ \hline
\end{tabular} 
\caption{Glueball mass in GeV extracted at different probing-momentum (in GeV) and at different instanton-size (in fm). In all calculations the diluteness has been kept fixed to $\kappa~\simeq~0.03$. The mass is only given for those values of $|{\bf p}|$ and $\bar \rho$ for which a plateau could be identified unambiguously.}
\end{table}

\section{The Glueball in the Instanton Vacuum}
\label{considerations}

In this section, we comment on the results of our  calculation and discuss at  the qualitative level the dynamical mechanism underlying the binding of scalar glueballs, in the instanton vacuum.

Our starting point consists in observing  that the statement of the existence of a bound-state in the spectrum can be formulated as a statement about the dynamics of quantum de-localization of a point-like excitation, in the Euclidean space. 
In order to see this, we recall that a necessary and sufficient condition for the binding of a scalar glueball in gluondynamics is that the  scalar point-to-point correlation function behaves as
\be
\Pi_S(\tau,{\bf x})\stackrel{\tau\to\infty}{=} \langle g^2 G^a_{\mu \nu}G^a_{\mu \nu}\rangle + 
\lambda_S^2~\frac{M_{0^{++}}}{4 \pi^2 \sqrt{\tau^2+{\bf x}^2}}
~K_1[~M_{0^{++}}\,\sqrt{\tau^2+{\bf x}^2}~],
\label{singleexpx}
\ee
where the first term is the  gluon condensate, while the second term is the propagator of a scalar particle of mass $M_{O^{++}}$ and $\lambda_S$ is the coupling of the scalar glueball state to the overlapping operator. 
\begin{figure}[t]
\begin{center}
\vspace{5mm}
\includegraphics[width=5.2cm,angle=0]{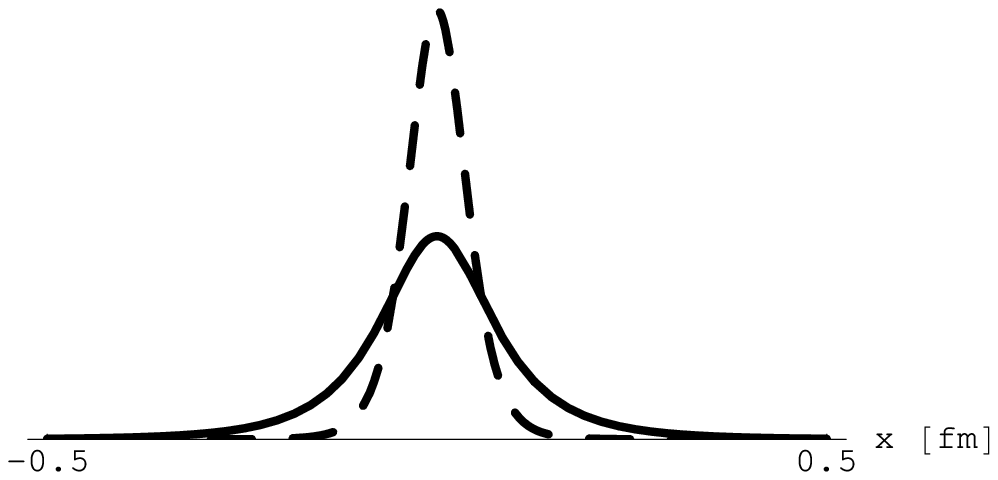}\hspace{0.2cm}
\includegraphics[width=5.2cm,angle=0]{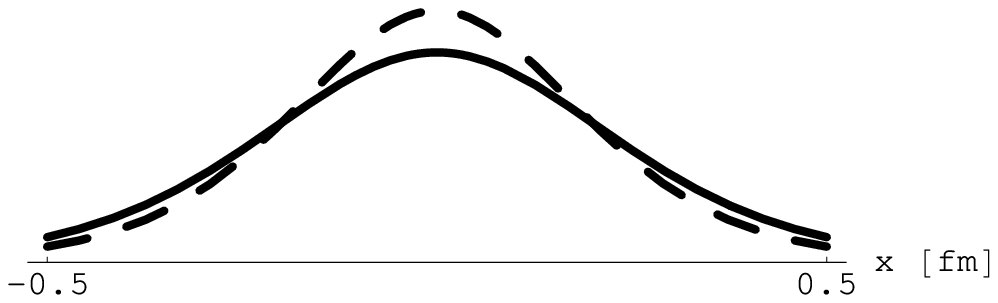}\hspace{0.2cm}
\includegraphics[width=5.2cm,angle=0]{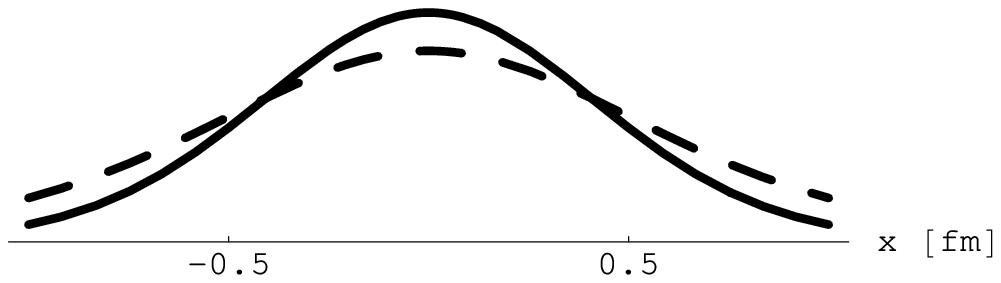}
\end{center}
\caption{The (normalized) scalar correlators in co-ordinate representation $\Pi'(\tau,|{\bf x}|)$, calculated at different Euclidean times: $\tau=0.1$~fm (left panel), $\tau=0.5$~fm (center panel) and $\tau=1.5$~fm (right panel). The dashed lines correspond to the propagators in the free gluon theory, while the solid lines represent the propagator of a bound-state of mass $M=1.5~$GeV. }
\label{compare}
\end{figure}

The corresponding two-point function in momentum-space representation, defined in Eq.(\ref{proj}) decays like a single exponential at large Euclidean times, for any fixed momentum ${\bf p}$,
\be
G_S(\tau,{\bf p})\stackrel{\tau\to\infty}{=} \lambda_S^2 \exp[- \,\tau ~\sqrt{{\bf p}^2+M_{0^{++}}^2}].
\label{singleexp}
\ee
We note that this equation implies that the correlator $G_S(\tau,{\bf p})$ will be dominated by lower and lower  momentum  components, as the Euclidean time increases. Hence, by uncertainty principle, Eq.s (\ref{singleexp}) and (\ref{singleexpx}) express the progressive de-localization of the center-of-mass of a free single-particle quantum state.  

This behavior differs strongly from the characteristics of the free gluon theory.
A gluonic excitation with $0^{++}$ quantum numbers created in the origin at time $\tau=0$, will diffuse according to the power-law
\be
\Pi^{free}_S(\tau,{\bf x})\stackrel{\tau\to\infty}{\sim} \frac{1}{(\tau^2+{\bf x}^2)^4}.
\label{freexlaw}
\ee 
In momentum-space representation, this corresponds  to the law
\be
G^{free}_S(\tau,{\bf p}) \stackrel{\tau\to\infty}{\sim} \frac{{\bf p}^2}{\tau^3}~\exp(-\tau |{\bf p}|).
\label{freeplaw}
\ee

The spatial dependence of the propagator in coordinate-space representation of a scalar excitation in the free theory and in a theory with a scalar glueball bound-state are confronted in Fig.\ref{compare}. 
We note that, for small Euclidean times, the free theory correlator is more peaked than in the case of bound-state correlator. On the other hand, at large Euclidean times, the situation is reversed: the free correlator is more spread-out than the single-particle correlator.  In other words, as expected,  free gluons de-localize more rapidly than a system of interacting gluons, bound in a glueball.

Let us now argue on the picture which emerges from our instanton model calculation for  gluon-dynamics. The vacuum of QCD and of Yang-Mills theory is characterized by the occurring of tunneling transitions, between degenerate classical vacua. During each of such transitions, gauge field fluctuations  with solitonic structure emerge: the instantons. Since such fields are quasi-classical, they represent collective  states made by a large number of gluons. On average, instanton fluctuations are isotropically distributed over space-time and hence give raise to a finite vacuum expectation value of the gluon field strength, i.e. the  gluon condensate. 
At vanishing (four-) momentum,  no further structure of the vacuum gauge field fluctuations is evident.

At finite momenta, however, the local structures of the instanton fluctuations begin to manifest themselves. For sufficiently short times, the short-distance correlations introduced by a single-instanton --- Eq.~(\ref{SIAcoord})--- dominate over the power-law correlations of the free-theory ---Eq.~(\ref{freexlaw})---. As shown in Eq.~(\ref{siacorr}), the occurrence of such  a tunneling transition  pumps-in additional high-momentum components into the momentum-projected correlator (\ref{freeplaw}). Our results have shown that such components are sufficient to locally turn the free-gluon de-localization law ---Eq.~(\ref{freeplaw})--- into  a single-particle de-localization law ---~Eq.~(\ref{singleexp})---. 

Such a dynamical correlation  can be pictured as following: during a quantum vacuum rearrangement,  gluons 
interact by exchanging momentum with the field of an instanton.  Obviously, a single tunneling event is not sufficient to correlate the field of a glueball for arbitrarily large Euclidean distances.
 In order to generate a state which diffuses like a single-particle state,  for arbitrary Euclidean times and arbitrarily low-momenta, one needs to re-sum the contribution of an infinite sequence of tunneling transitions, i.e. to account for an infinite number of scattering processes of the gluons with the instanton fields. 

\section{Conclusions and outlook}
\label{conclusions}

In this work, we have studied the short-distance instanton-induced interactions in the scalar glueball channel.
In particular, we have addressed the question whether such interactions  are 
sufficient  to generate a glueball bound-state with $0^{++}$ quantum numbers and realistic mass.

We performed a SIA calculation in the quenched approximation, which in this channel corresponds to  gluon-dynamics.  We observed  evidence for a bound-state and we have estimated its  mass to be 
$M^{SIA}_{0^{++}}\simeq 1.290-1.420$~GeV, i.e. consistent with the value calculated by Meyer and Teper in lattice gauge theory.
We stress the fact that our results do not rely on any {\it a priori} assumption on the structure of the spectral function. The emergence of a plateau in the effective mass plot in this model 
represents a strong indication that the origin of the scalar glueball is mostly provided by short-sized non-perturbative gluon fluctuations, rather than by confining forces.

\acknowledgments
We thank H.B.~Meyer, J.W.~Negele and V.Vento for important discussions. 
This work was supported in part by the "Bruno Rossi" INFN/MIT exchange program.

\appendix
\section{Interacting Instanton Liquid Model}

In the following,  we shall illustrate a particular form of the one- and two-body correlations $G_1(\Omega_1)$ and $G_2(\Omega_1,\omega_2)$, which  appear in Eq.~(\ref{G1G2}). This choice defines  the 
so-called Interacting Instanton Liquid Model \cite{iilmfirst}. 
\begin{itemize}
\item The one-body term $G_1(\Omega)$, is derived from the quantum fluctuations around the single instanton configuration is calculated in Gaussian approximation, through 't~Hooft's semi-classical instanton amplitude $d(\rho_i)$\cite{thooft}
\begin{equation}
G_1\propto -\ln \left[\rho^{-5} \rho^{N_f} \beta_1(\rho)^{2N_c} \mbox{exp}\left(-\beta_2(\rho) +\left(2 N_c +\frac{b'}{2b}\right) \frac{b'}{2 b \beta_1(\rho)}\ln(\beta_1(\rho)) \right)\right]. \label{thooft}
\end{equation}
The functions $\beta_1$ and $\beta_2$ are defined as
\begin{eqnarray}
\beta_1(\rho) = - b \ln(\rho \Lambda_{PV}) \qquad
\beta_2(\rho) = \beta_1(\rho) +\frac{b'}{2 b} \ln \left( \frac 2 b \beta_1(\rho)\right),\nonumber
\end{eqnarray}
where $\Lambda_{PV}=1.03~\Lambda_{\overline{MS}}$ is the QCD scale in the Pauli-Villars scheme. The coefficients $b, b'$ are given by
$b = \frac{11}{3}N_c - \frac 2 3 N_f$ and 
$b' = \frac{34}{3}N_c^2 - \frac{13}{3} N_c N_f + \frac{N_f}{N_c}$.
\item 
A classical two-body interaction between instanton and anti-instanton $G_{2}^{cl.}$ is derived from the Yang-Mills action evaluated in the stream-line {\it ansatz} (\ref{stream}),
\be
G_{2}^{cl.}(\Omega_i,\Omega_j) = S_{YM}[A_{\mu}^{stream.}(\Omega_i, \Omega_j)]- 2 S_0,
\ee
where $S_0= \frac{8 \pi}{g^2}$ is the action  of an isolated instanton.
The total stream-line gauge field is constructed by summing-up the field of instanton-antiinstanton pairs. The analytic gauge potential for a single  instanton-antiinstanton pair  is given by
\begin{equation}
	A_{\mu}^a= 2\eta_{\mu\nu}^a\frac{x_{\nu}}{x^2+\rho^2\lambda}+2R^{ab}\eta_{\mu\nu}^b\frac{\rho^2}{\lambda}\frac{1}{x^2(x^2+\rho^2/\lambda)}
	\label{stream}
\end{equation}	
where $\lambda$ is the streamline conformal parameter defined by 
\begin{eqnarray}
\lambda &=& \frac{R^2+\rho_I^2+\rho_A^2}{2\rho_I\rho_A}+\Big(\frac{(R^2+\rho_I^2+\rho_A^2)^2}{4\rho_I^2\rho_A^2}-1\Big)^{1/2}\\\nonumber
R &=& |z_I-z_A|
\end{eqnarray}
and 
\begin{equation}
	R^{ab}=\frac{1}{2}\textrm{Tr}[U^{\dagger}\tau^aU\tau^b]
\end{equation}
represents the relative color orientation of the two instantons.
\item
The effect of quantum two-body correlations is accounted for by means of a phenomenological short-distance hardcore term \cite{iilmfirst}:
\begin{equation}
\label{core}
	G^{phen.}_{2}(\Omega_i,\Omega_j)=\frac{A}{\lambda^4}|u|^2,
\end{equation}
where $u$ is a color orientational factor. Such a term removes large-sized instantons from the vacuum and provides a cut-off to the momentum that can be exchanged through the instanton field. Hence, it restricts the region of applicability of the model to the non-perturbative sector, characterized by momenta of the order  $p< 1/\bar{\rho}$, where $\bar{\rho}$ is the average instanton size.
The dimensionless coefficient $A$ in (\ref{core}) controls the strength of the repulsion and is the only phenomenological parameter of the model. Most calculations adopt the value $A\simeq128$, first suggested by Sch\"afer and Shuryak~\cite{iilmfirst} and used also in \cite{nucleonILM, resonancesILM}. 
\end{itemize}


\begin{thebibliography}{99}
\bibitem{pheno} F.E.Close, Int. J. Mod. Phys. {\bf A20} (2005), 5156. 
M.R.Pennington, J. Phys. Conf. Ser. {\bf 18} (2005),1. 
\bibitem{gluex} http://dustbunny.physics.indiana.edu/HallD/
\bibitem{Bettoni:2005ut}
  D.~Bettoni, 
  J.\ Phys.\ Conf.\ Ser.\  {\bf 9} (2005) 309.
\bibitem{liu} 
 C.J.Morningstar and M.Peardon, Phys. Rev. {\bf D60} (1999), 034509. 
  Y.~Chen {\it et al.},
  Phys.\ Rev.\  D {\bf 73} (2006), 014516
\bibitem{mayer} 
  H.~B.~Meyer and M.~J.~Teper,
  Phys.\ Lett.\  B {\bf 605} (2005), 344 
H. B. Meyer, "Glueball Regge Trajectories", Ph.D. Thesis, Oxford University (unpublished) [arXiv:hep-lat/0508002].
\bibitem{sizeglue} F. de Forcand and K-F Liu, Phys. Rev. Lett. {\bf 69} (1992) 245.
\bibitem{shuryak82} E.V.Shuryak,  Nucl. Phys. {\bf B214} (1982), 237. 
\bibitem{diakonov84} D.~Diakonov and V.~Petrov: Nucl. Phys. {\bf B245} (1984), 259.
\bibitem{gattringer} C. Gattringer, Phys. Rev. Lett. {\bf 88} (2002), 221601.
C. Gattringer {\it et al.}, Nucl. Phys. {\bf B617} (2001) 101.
\bibitem{faccioli} P. Faccioli and T. A. De Grand, Phys. Rev. Lett. {\bf 91} (2003), 182001.
\bibitem{nucleonILM} M.Cristoforetti, P.Faccioli, J.Negele and M.Traini,
Phys. Rev. {\bf D75} (2007), 034008. 
\bibitem{mymasses} P.Faccioli, Phys. Rev. {\bf D65} (2002) 094014.
\bibitem{ILM3pt} P. Faccioli and E.V. Shuryak,  Phys. Rev. {\bf D65} (2002) 076002.
\bibitem{nucleonFF}  P. Faccioli, Phys. Rev. {\bf C69} (2004) 065211. 
P. Faccioli and E.V. Shuryak,  Phys. Rev. {\bf D65} (2002) 076002.
 P. Faccioli, A. Schwenk and E.V. Shuryak, Phys. Lett. \textbf{B549} 
(2002) 93
\bibitem{pionFF}  P. Faccioli, A. Schwenk and E.V. Shuryak, Phys. Rev. \textbf{D67} (2003) 113009.
\bibitem{delta12} 
M. Cristoforetti {\it et al.} Phys. Rev. {\bf D70} (2004) 054016
\bibitem{resonancesILM} M.Cristoforetti, P.Faccioli and M.Traini
Phys. Rev. {\bf D75} (2007), 054024.
\bibitem{earlyOPE1} V.A. Novikov, M.A.Shifman, A.I.Vainsthein and V.I. Zakharov, Nucl. Phys. {\bf B191} (1981), 301. 
\bibitem{earlyOPE2} E.V. Shuryak, Nucl. Phys. {\bf B203} (1982) 116.
\bibitem{forkel} H.Forkel, Phys. Rev. {\bf D64} (2001), 034015. H. Forkel, Phys. Rev. {\bf D71} (2005), 054008.
\bibitem{shusha} T. Schafer and E.V.Shuryak, Phys. Rev. Lett. {\bf 75} (1995) 1707.
\bibitem{shuryakrev} T.~Sch\"afer and E.V.~Shuryak: Rev. Mod. Phys. {\bf 70} (1998), 323.
\bibitem{thooft} G. 't~Hooft, Phys. Rev. Lett. \textbf{37} (1976) 8.
G. 't~Hooft, Phys. Rev. \textbf{D14} (1976) 3432.
\bibitem{bnumber} P.Faccioli,  Phys. Rev. {\bf D71} (Rapid Comm.) (2005) 091502 
\bibitem{EDM} P.~Faccioli, D.~Guadagnoli and S.~Simula, Phys. Rev. {\bf D 70} (2004) 074017 
\bibitem{sia} P. Faccioli and E.V. Shuryak,  Phys. Rev. {\bf D64} (2002) 114020.
\bibitem{streamline} J.J.M. Verbaarshot, Nucl. Phys. {\bf B362} (1991) 33.
\bibitem{iilmfirst} T.Sch\"afer and E.V.Shuryak, Phys. Rev. {\bf D53} (1996), 6522. T.Sch\"afer and E.V.Shuryak, Phys. Rev. {\bf D54} (1996), 1099.
\bibitem{chu} Chu {\it et al.} Phys. Rev. {\bf D49} (1994) 6039.
\bibitem{latticerho} J.W. Negele, Nucl. Phys. Proc. Suppl. \textbf{73} (1999) 92.
\bibitem{SIAglue} Novikov {\it et al.} Nucl. Phys. {\bf B165} (1979), 67.  E.V. Shuryak, Nucl. Phys. {\bf B203} (1982) 93.
\end{thebibliography}
\end{document}